\documentclass[11pt,twoside]{article}
\usepackage{newpasp}
\usepackage{epsf}
\begin{document}
\title{Hot Molecular Cores and High-Mass Star Formation}
\author{F.F.S.\ van der Tak}
\affil{Max-Planck-Institut f\"ur Radioastronomie, Auf dem H\"ugel 69,
 53121 Bonn, Germany}

\begin{abstract}
  This review covers hot cores in the context of high-mass star
  formation. After giving an overview of chemical processes and
  diversity during high-mass star formation, it reviews the `warm
  envelope' phase which probably precedes the formation of hot cores.
  Some recent determinations of the cosmic-ray ionization rate are
  discussed, as well as recent evidence for hot cores around low-mass
  stars. Routes for future hot core research are outlined.
\end{abstract}

\section{Introduction}
\label{intro}

Named after the Orion Hot Core, hot molecular cores are small
($d\la$0.1~pc) pockets of warm ($\ga 100$~K) dense ($\ga
10^6$~cm$^{-3}$) molecular gas, which show very rich submillimeter
spectra.  This definition includes both internally and externally heated
clumps near stars of all masses, but the focus of this review is on
the case of internal heating by young high-mass stars, which form the
majority of known hot cores. This review only outlines basic
properties of hot cores with an emphasis on recent developments.
Recent reviews on the chemistry of star-forming regions, including hot
cores, have been given by van Dishoeck \& van der Tak (2000) and Langer et
al.\ (2000); high-mass star formation is discussed by Kurtz et al.\ 
(2000), Churchwell (2002) and Garay (this volume).

\section{The embedded phase of high-mass star formation}
\label{hmsf}

Stars of masses $>$8~M$_\odot$ spend a significant fraction of their
lifetime, $>$10\%, embedded in their natal molecular clouds. This
embedded phase can be subdivided into four observationally different
groups of objects: \textbf{(1)} The initial conditions of high-mass
star formation should be large mass reservoirs which are local
temperature minima and density maxima.  Candidate objects in this
phase have been found through infrared (Egan et al.\ 1998; Carey et
al.\ 1998) and submillimeter (Motte et al.\ 2003) observations.
\textbf{(2)} High-mass protostellar objects, where the central star is
surrounded by a massive envelope with a centrally peaked temperature
distribution (van der Tak et al.\ 2000a, Sridharan et al.\ 2002).
Close to the star, ices start evaporating off the grains.
\textbf{(3)} Hot cores, with large masses of warm and dense molecular
gas, and large abundances of complex organic molecules such as
CH$_3$OCH$_3$ and CH$_3$OCHO. These species are thought to be the
chemical `daughter' products of reactions of evaporated ice mantle
components, their `parents'.  \textbf{(4)} Small ($d\la$0.1~pc)
pockets of ionized gas, called ultracompact H~{\sc ii} regions. The
ordering of these phases from `cold' to `hot' may well be an
evolutionary sequence, where the ratio of envelope mass to luminosity
decreases, although the mass of the original cloud may also play a
role.  Phases 2 and 3 are accompanied by massive molecular outflows
(Beuther et al.\ 2002), phases 1--3 by CH$_3$OH masers (Walsh et al.\ 
2001), and H$_2$O masers can be found in all four phases.

The separation between these four phases is not as clean as in the
case of low mass star formation. Massive stars form in groups and the
four above phases often occur right next to each other, making it hard
to assign a single age to the region.  An example is G10.47, where VLA
images show peaks in both the NH$_3$ (4,4) line, which traces
warm molecular gas, and in the 1.3~cm continuum, which traces ionized
gas (Cesaroni et al.\ 1998).  In W49, Wilner et al.\ (2001) imaged
CH$_3$CN line and dust continuum emission at 1.4~mm using the BIMA
interferometer. The number of hot cores in W49 is about half that of
ultracompact H~{\sc ii} regions, indicating that the hot core phase
lasts $\sim$half the lifetimes of ultracompact H~{\sc ii} regions of
$\sim$10$^5$~yr.  This value agrees well with estimates based on
chemical modeling (e.g., Rodgers \& Charnley 2001) and with dynamical
timescales of molecular outflows. This agreement, together with the
lack of spatial ordering of different phases within the region, argues
against triggering of star formation by an external event. 

\section{Chemical processes during high-mass star formation}
\label{chem}

In the earliest, dense, cold phases prior to star formation, atoms and
molecules freeze out onto grain surfaces, which act as catalysts for
neutral-neutral reactions. The composition of the resulting grain
mantles can be probed by mid-infrared spectroscopy. This technique has
developed greatly with the ISO mission and the advent of high-quality
spectrographs on 8m-class telescopes. The derived abundances
(Table~\ref{tab:ice_comp}) exceed by orders of magnitude the values
that can be produced by reactions in the gas phase, with CO as the only
exception. The data thus make a strong case for an active grain
surface chemistry.  Additional evidence comes from the observed
isotopic composition (D/H ratio) of evaporated H$_2$CO and CH$_3$OH
(Charnley et al.\ 1997).

However, the detailed reaction mechanisms of surface chemistry are not
well understood. Laboratory experiments on hydrogenation of CO give
conflicting results (Watanabe \& Kouchi 2002; Hiraoka et al.\ 2002),
so that the origin of solid CH$_3$OH is not clear. Also, observational
limits on solid HDO are inconsistent with the predictions of surface
chemistry models (Dartois et al.\ 2003).

Considering elemental abundances, it is clear that the observational
inventory of the solid state is incomplete. The likely reason is that
the major carriers of some elements are undetectable through infrared
spectroscopy. For example, the only N-bearing species is NH$_3$, which
contains only $\approx$5\% of elemental nitrogen. The remaining
nitrogen could be in the (undetectable) form of N$_2$. Similarly, OCS
does not make up the full sulphur budget. Submillimeter observations
suggest trace amounts ($\la$1\%) of solid H$_2$S or SO$_2$
evaporating, but observations with ISO rule out both species, as well
as atomic S, as major sulphur carriers (van der Tak et al.\ 2003).

\begin{table}[t]
  \caption{Composition of grain mantles in star-forming regions}
  \label{tab:ice_comp}
  \begin{center}
  \begin{tabular}{ccc}
\noalign{\bigskip}
\tableline
\noalign{\medskip}
Species & Abundance & Reference \\
\noalign{\medskip}
\tableline
\noalign{\medskip}
H$_2$O   & 10$^{-4}$ of H$_2$ & Van Dishoeck 1998  \\    
NH$_3$   & $\la$7\% of H$_2$O & Dartois et al.\ 2002  \\ 
H$_2$CO  & 3--6\% of H$_2$O   & Gibb et al.\ 2000     \\ 
CH$_3$OH & 3--30\% of H$_2$O  & Dartois et al.\ 1999  \\ 
CO$_2$   & 14--20\% of H$_2$O & Gerakines et al.\ 1999 \\
CO       & 2--25\% of H$_2$O  & Tielens et al.\ 1991 \\
OCS      & 0.1\% of H$_2$O    & Palumbo et al.\ 1997 \\
CH$_4$   & 2--4\% of H$_2$O   & Boogert et al.\ 1998 \\
HCOOH    & 3--7\% of H$_2$O   & Schutte et al.\ 1999 \\
\noalign{\medskip}
\tableline
  \end{tabular}
  \end{center}

\end{table}

After evaporation into the gas phase, the molecules of
Table~\ref{tab:ice_comp} start a complex reaction scheme, building up
long carbon chains with many different functional groups.
Spectroscopy of hot cores at millimeter wavelengths has revealed
considerable chemical complexity: the BIMA survey of CH$_3$COOH by
Remijan et al.\ (2003) provides a recent example.  Probably even
larger molecules are formed, but these are very hard to detect,
because of spectroscopic confusion, and because individual spectral
lines become weaker as the partition function increases for ever
larger species.

The formation of large molecules in hot cores is mostly due to NH$_3$
and CH$_3$OH, which, after acquiring a proton from H$_3^+$, H$_3$O$^+$
or HCO$^+$, can form chains.  The HCN and C$_2$H$_5$OH
molecules could play similar roles, but so far, only upper limits on
their solid state abundances have been obtained.  An important
determinant of the chemistry is therefore the competition for protons
between NH$_3$ and CH$_3$OH, which both have a high proton affinity.
Rodgers \& Charnley (2001) show model results for various values of
the NH$_3$/CH$_3$OH ratio, which observations do not yet constrain
well. 

Chemical diversity can also be the result of
high-temperature neutral-neutral reactions.  At temperatures
$\ga$300~K, the reactions of O and OH with H$_2$ drive all the
available oxygen into H$_2$O.  Cores at high temperatures are
therefore expected to be rich in nitrogen-bearing compounds (Rodgers
\& Charnley 2001). Although in the models, age differences also cause
chemical diversity, observations seem to support a link between
temperature and nitrogen-oxygen differentiation. For example, IRAM
Plateau de Bure data show that the peak position of nitrogen-bearing
species in W3 (OH) has a higher temperature than that of
oxygen-bearing species (Wyrowski et al.\ 1999). Additional support
comes from JCMT spectroscopy of the HCN $J=9\to8$ line at 797~GHz
(Boonman et al.\ 2001). The data show that the HCN abundance in the
envelope of the massive young star AFGL 2591 is not constant with
radius, but is enhanced by a factor of 100 at small radii where
$T\ga300$~K.

Shocks do not seem as important for the chemistry of high-mass
star-forming regions as they are in the low-mass case (e.g., Bachiller
et al.\ 2001).  Still, two molecules suggest that the envelopes of
young high-mass stars may have been processed by shocks, although
neither one conclusively. First, ISO-SWS data show low ratios of
gas-phase to solid-state CO$_2$ ($\la 0.1$) and the CO$_2$ gas phase
abundance remains low through the 100--300~K temperature regime
(Boonman et al.\ 2003).  After evaporating off grains, CO$_2$ must be
promptly destroyed, which shocks can do in reactions with H (Charnley
\& Kaufman 2000) or perhaps H$_2$ (Doty et al.\ 2002; Talbi \& Herbst
2002).  Second, Hatchell \& Viti (2002) measured NS/CS ratios of
0.02-0.05 in a sample of six hot cores and interpreted these as
evidence for shocks. The main reactions to form NS require SH and NH
which are produced from OH.  Shocks use OH to form H$_2$O and suppress
the production of NS.  The values of NS/CS=0.001 -- 0.01 measured in
warm envelopes (van der Tak et al.\ 2003) may indicate that shocks
play a role too. However, for $t=3\times 10^4$~yr, the Doty et al.\ 
(2002) model of envelope chemistry also predicts NS/CS $\sim 10^{-3}$,
so this ratio cannot be used to demonstrate the influence of shocks.

\section{Hot cores with cool envelopes}
\label{envelope}


Before the onset of an ultracompact H~{\sc ii} region, young high-mass
stars are surrounded by envelopes where ice evaporation occurs in the
inner parts. Time-dependent chemical models for this phase were
constructed by Doty et al.\ (2002). These models employ standard
reaction networks, and the initial conditions resemble those of dark
clouds in the outer envelope ($T<100$~K). In the inner envelope
($T>100$~K), the initial abundances of the species in
Table~\ref{tab:ice_comp} are enhanced, corresponding to evaporation of
ice mantles.

\begin{figure}[t]
  \plotfiddle{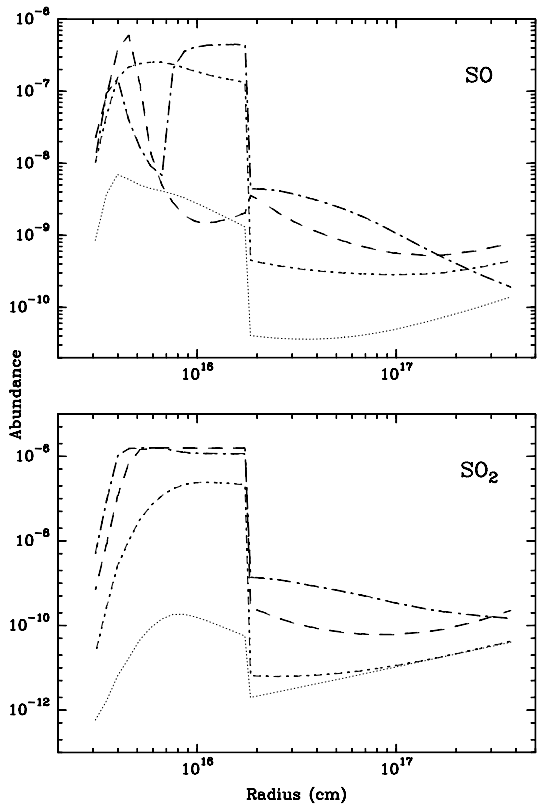}{10cm}{0}{40}{40}{-150}{-20} 

  \caption{Abundance profiles of gas-phase SO and SO$_2$ predicted for the
    envelope of the massive protostar AFGL 2591 for chemical ages of
    300, 3000, 30000 and 300000~yr (curves from bottom to top). After
    Doty et al.\ 2002. }

  \label{fig:doty}
\end{figure}

Figure~\ref{fig:doty} shows results for SO and SO$_2$ as an example.
The evaporation of ice mantles leads to pronounced abundance
enhancements (`jumps') for several molecules.  Observational evidence
for such jumps is mounting: H$_2$CO, CH$_3$OH, SO, SO$_2$ and OCS show
jumps at $T\approx 100$~K (van der Tak et al.\ 2000b; Sch\"oier et
al.\ 2002; van der Tak et al.\ 2003), likely due to ice evaporation.
In contrast, HCN is enhanced at $T\approx 300$~K (Boonman et al.\ 
2001) in the source AFGL 2591. Such an enhancement is a predicted
effect of the O$\to$H$_2$O reaction and suggests that in this source,
high-temperature gas-phase reactions have started, marking the birth
of the hot core phase.  However, the abundance jumps have so far been
inferred from detailed modeling of single-dish spectra.
Millimeter-wave interferometers could resolve the jumps spatially,
which would be an important test of current models of envelope
chemistry.

A subset of these `warm envelopes', where the high mid-infrared
brightness allows detection of H$_3^+$ rovibrational absorption lines,
has been used to derive the cosmic-ray ionization rate (van der Tak \&
van Dishoeck 2000).\footnote{In this paper, the values in the last
  column of Table~1 were scaled the wrong way around. The conclusions
  of the paper are not affected though.}  This process is the main
ionization mechanism of dense molecular clouds, regulating for example
the influence of magnetic fields on their dynamics. Van der Tak \& van
Dishoeck compared observations of H$_3^+$ absorption and
H$^{13}$CO$^+$ emission to a small chemical network coupled to
detailed models of the sources' structure, to find a mean $\zeta$ of
$2.6\times 10^{-17}$~s$^{-1}$. The source-to-source spread of a factor
of 2--3 may be correlated with $N$(H$_2$), suggesting that absorption
of cosmic rays plays a role. The value of $\zeta=5.6\times
10^{-17}$~s$^{-1}$ for one particular source was verified by Doty et
al.\ (2002) using a detailed chemical model.\footnote{In this
  paper, the predicted $N$(H$_3^+$) should read $5\times
  10^{13}$~cm$^{-2}$ rather than $5\times 10^{14}$~cm$^{-2}$.} 

McCall et al.\ (2003) infer a much higher value of $\zeta\approx 1
\times 10^{15}$~s$^{-1}$ for the diffuse cloud towards $\zeta$~Per. More
detailed modeling of the chemistry of H$_3^+$ and OH and HD, which
also depend on $\zeta$, give similar results within a factor of 2--3,
$\sim$10$\times$ higher than the results for the massive protostars.
McCall et al.\ propose low-energy cosmic rays that penetrate diffuse
but not dense clouds. High-resolution observations of HD in a large
sample of diffuse clouds with FUSE will help to test this conclusion.


A much lower value of $\zeta=0.6\times 10^{-17}$~s$^{-1}$ was obtained
by Caselli et al.\ (2002) for the dense core of L1544. This result
needs to be checked in view of the large depletion of CNO-bearing
molecules implied by the large abundance of H$_2$D$^+$ towards this
source (Caselli et al.\ 2003). If confirmed, this low value of $\zeta$
provides additional evidence for variations in the ionization rates of
molecular clouds caused by absorption of cosmic rays. However, more
sources need to be studied before local variations in cosmic-ray
production can be ruled out.

\section{Low-mass hot cores}
\label{corinos}

Although the existence of warm ($>$100~K) gas in the inner envelopes
of low-mass protostars was well-known, it was unclear whether such
objects achieved the same chemical complexity as their high-mass
counterparts (e.g., Sch\"oier et al.\ 2002). This situation changed
when Cazaux et al.\ (2003) used the IRAM 30m telescope to detect both
'parent' and 'daughter' type species in the envelope of the low-mass
protostar IRAS 16293--2422. The abundances are similar to or perhaps
even higher than those in the Orion hot core; the assumed $N$(H$_2$)
of the inner envelope of IRAS 16293 and the size of the Orion hot core
are critical to this comparison.  The high deuterium fractionation of
CH$_3$OH measured by Parise et al.\ (2002) for IRAS 16293 may also be
reproduced by grain surface chemistry models if the atomic D/H ratios
from the model of Roberts et al.\ (2003) are adopted. However, the
impact of the rate coefficients of key H$\leftrightarrow$D reactions
measured by Gerlich et al.\ (2002) on the Roberts et al.\ model needs
to be studied.

The initial chemical conditions of low- and high-mass hot cores may
well be similar. Using ISAAC on the VLT, Pontoppidan et al.\ (2003)
detected CH$_3$OH ice in three low-mass young stellar objects. The
solid CH$_3$OH abundances of 15--25\% relative to H$_2$O are
comparable to those in the most methanol-rich massive sources.
However, it is unclear if evaporation of these ice mantles can drive
the rich gas-phase chemistry of IRAS 16293. The measured collapse
motions in the envelope imply a travel time for gas through the hot
core region of only a few hundred years, which is too short to build
up high abundances of complex molecules (Sch\"oier et al.\ 2002;
Rodgers \& Charnley 2003).

\section{Future directions}
\label{future}

Interferometry at (sub)mm wavelengths allows imaging of dust continuum
and molecular lines at 1$''$ or slightly higher resolution.  The
contribution by Beuther in this volume shows a recent example obtained
with the Sub Millimeter Array.  Such data are useful to constrain the
temperature and density structure of high-mass star-forming cores, and
to determine the multiplicity of their power sources (Wyrowski et al.\ 
1999). However, current instruments can spatially resolve such
structure only in a few nearby objects, and searches for kinematic
structure in hot cores have just started (e.g., Cesaroni et al.\ 
1998).  Efforts are also hampered by limited sensitivity which at the
moment prohibits observation of multiple transitions and of isotopic
species, which is necessary to determine the optical depth and the
excitation of the lines.  In the future,
ALMA 
will provide
increases in angular resolution and sensitivity by factors of 10 or
more.

Observational progress will also come from the
HIFI 
instrument onboard
Herschel, which will perform spectral line surveys at heterodyne
resolution.  Unhindered by the Earth's atmosphere, HIFI will not only
probe H$_2$O (and maybe O$_2$) directly, but also observe lines that
are blocked by atmospheric H$_2$O (especially the ground state
transitions of hydride molecules). The calibration will be much better
than can be achieved from the ground. However, the success of the HIFI
mission depends on the availability of spectroscopic and collisional
data for astrophysically interesting molecules at frequencies up to
2~THz. Sideband deconvolution is another critical issue (e.g., Comito
\& Schilke 2002).  These caveats are especially important for hot
cores with their rich line spectra.


\acknowledgments The author thanks Fredrik Sch\"oier, Friedrich
Wyrowski and Ewine van Dishoeck for useful discussions.

\end{document}